# Multiple Embedded Eigenstates in Nonlocal Plasmonic Nanostructures

Solange V. Silva[1], Tiago A. Morgado[1], Mário G. Silveirinha[1,2*]

[1]*Instituto de Telecomunicações and Department of Electrical Engineering, University of Coimbra, 3030-290 Coimbra, Portugal*

[2]*University of Lisbon, Instituto Superior Técnico, Avenida Rovisco Pais, 1, 1049-001 Lisboa, Portugal*

E-mail: solange@co.it.pt, tiago.morgado@co.it.pt, mario.silveirinha@co.it.pt

## Abstract

Trapping light in open cavities is a long sought "holy grail" of nanophotonics. Plasmonic materials may offer a unique opportunity in this context, as they may fully suppress the radiation loss and enable the observation of spatially localized light states with infinite lifetime in an open system. Here, we investigate how the spatial dispersion effects, e.g., caused by the electron-electron interactions in a metal, affect the trapped eigenstates. Heuristically, one may expect that the repulsive-type electron-electron interactions should act against light localization, and thereby that they should have a negative impact on the formation of the embedded eigenstates. Surprisingly, here we find that the nonlocality of the material response creates new degrees of freedom and relaxes the requirements for the observation of trapped light. In particular, a zero-permittivity condition is no longer mandatory and the same resonator shell can potentially suppress the radiation loss at multiple frequencies.

---

[*] To whom correspondence should be addressed: E-mail: mario.silveirinha@co.it.pt

# Main Text

Confining and storing light into a tiny volume for many oscillation periods is a long sought "holy grail" in photonics [1]. This goal remains elusive today, because the coupling with the surrounding environment invariably leads to absorption and radiation losses. The standard way to confine light into some space-region is by using opaque physical barriers, e.g., reflecting mirrors [2] or photonic band-gap materials [3,4], or alternatively by exploiting the total internal reflection as in whispering gallery resonators [5,6]. However, these resonators typically need to have dimensions much larger than the radiation wavelength to effectively block the radiation leakage to the exterior. Other mechanisms that promote the light localization rely on plasmonics [7-8], weakly radiating anapole modes in dielectric nanoparticles [9] and Anderson localization [10,11].

Over the last decade, it was shown that optical bound states with suppressed radiation loss may exist within the radiation continuum in *open* resonators [12-22]. These excitations are known as "embedded eigenstates", in analogy with the spatially localized electron states with "positive energy" in quantum systems [23-25]. Importantly, for typical structures, e.g., dielectric photonic crystal based designs [12-16], the radiation loss can be fully suppressed only if the open resonator is spatially unbounded. If the open resonator is spatially truncated the lifetime of the embedded eigenstate becomes finite, even in the ideal case of vanishing material absorption [26].

Remarkably, it was proven in [26] that plasmonic materials offer a unique opportunity to totally suppress the radiation loss in open and spatially-bounded three-dimensional (3D) nanostructures [26]. This idea was further explored in subsequent works [27-30]. In particular, a core-shell nanoparticle formed by a dielectric core and an epsilon-near-zero (ENZ) shell may be designed to support wave oscillations with infinite lifetime in the limit of vanishing material loss [26]. The same mechanism can be used to confine light in arbitrarily shaped optical cavities [28, 30]. The emergence of the embedded eigenstates is rooted in the excitation of volume plasmons in the ENZ shell, which effectively prevent the wave in the core from escaping to the exterior.

The studies of Refs. [26-30] assumed that the plasmonic ENZ shell had a local response, i.e., the material permittivity was assumed independent of the spatial variation of the fields. Possible effects of spatial dispersion were only superficially discussed in [26]. In metals the nonlocal effects arise primarily due to many-body electron-electron (repulsive-type) interactions, and are usually modelled through a *diffusion*-term in the framework of the hydrodynamic model [31-41]. Nonlocal effects may be critically important in plasmonics, especially for nanosized particles [31-43]. Thus, one might think that spatial dispersion would be an additional obstacle to create embedded eigenstates. Surprisingly, we prove in this Letter that it is precisely the opposite, and that nonlocal effects offer a unique path to localize light in an open resonator. It is shown that the conditions for the observation of embedded eigenstates are very much relaxed when nonlocal effects are taken into account; in particular, the shell permittivity is not anymore constrained to be precisely zero.

Figure 1(a) illustrates the geometry of the core-shell meta-atom. It consists of a bi-layered spherical nanoparticle standing in air. The core region and the outer shell have radii $R_1$ and $R_2$, respectively. The core material is a simple dielectric with relative permittivity $\varepsilon_1$, e.g., air, and the shell is made of a plasmonic material, e.g., a noble or alkali metal at optical frequencies or a semiconductor in the terahertz regime. We use the hydrodynamic model [37] to model the nonlocal effects in the shell. The unbounded plasmonic material supports three plane-wave modes with a spatial dependence of the type $e^{i\mathbf{k}\cdot\mathbf{r}}$: two transverse waves and also a longitudinal wave [37,39]. The relative permittivity seen by the transverse modes is described by a standard Drude dispersion model $\varepsilon_{2,T}(\omega) = \varepsilon_\infty - \omega_p^2 / \left[\omega(\omega + i\omega_c)\right]$, where $\omega_p$ is the plasma frequency, $\omega_c$ is the collision frequency, and $\varepsilon_\infty$ is the high-frequency relative permittivity. On the other hand, the longitudinal mode is described by the wave vector dependent relative permittivity $\varepsilon_{2,L}(\mathbf{k},\omega) = \varepsilon_\infty - \omega_p^2 / \left[ \omega(\omega + i\omega_c) - \beta^2 k^2 \right]$ [37], where $\beta^2 = 3/5 v_F^2$ and $v_F$ is the Fermi velocity [35-36]. The nonlocality strength parameter $\beta/c$ may reach values on the order of $1/450$ in alkali metals [31], $1/280$ in semiconductors [41], and even larger values in metamaterials [44, 45]. Without loss of generality, it is assumed throughout this manuscript that $\varepsilon_\infty = 1$.

Due to the spherical symmetry, the natural modes of the core-shell nanoparticle (Fig. 1(a)) can be split into transverse radial magnetic ($TM^r$) and transverse radial electric ($TE^r$) waves. We focus on the $TM^r$ modes whose properties are determined by the hybridization of transverse and longitudinal waves. Using Mie theory [46-47] the electromagnetic fields may be written in all the regions of space in terms of spherical Bessel functions [26]. The fields in the different regions are linked through the standard boundary conditions of the hydrodynamic model. In this manner, the modal problem is reduced to a $6 \times 6$ homogeneous linear system of the form $\mathbf{M} \cdot \mathbf{x} = 0$; for more details see the Supplemental Material [48]. The oscillation frequencies $\omega = \omega' + i\omega''$ ($\omega'' \leq 0$) of the $TM^r$ modes of oscillation are given by the nontrivial solutions of the characteristic equation $D(\omega, R_1, R_2, \varepsilon_1, \omega_p, \beta) \equiv \det(\mathbf{M}) = 0$.

In the local limit ($\beta = 0$), i.e., for an electron gas with non-interacting electrons, it is known from Ref. [26] that the embedded eigenstates can occur only if the shell has a zero-permittivity, i.e., $\varepsilon_2 = 0$ is a mandatory condition. Thus, the oscillation frequency of a trapped state is necessarily $\omega = \omega_p$. An $\varepsilon_2 = 0$ shell behaves as a perfect magnetic (PMC) wall for $TM^r$ waves. The embedded eigenstates are formed when $\omega = \omega_p$ coincides with an eigenfrequency of the equivalent PMC resonator, i.e., the core surrounded by a fictitious

PMC boundary. For an embedded eigenstate with dipolar-type symmetry, this condition leads to the geometrical constraint $R_1 = R_{1,0} \equiv 4.49c/\left(\omega_p \sqrt{\varepsilon_1}\right)$ [26].

In general, for a spatially dispersive shell, the embedded eigenstates are solutions of $D = 0$ with a real-valued $\omega$ so that the oscillations do not decay with time. The simplest way to understand the general structure of the solutions and to generate them is by using a reduced dispersion equation, which is obtained as follows. For an embedded eigenstate the fields in the air region ($r > R_2$) are required to be identically zero [26]; hence the tangential electric and magnetic fields evaluated inside the shell must vanish at $r = R_2^-$. In addition, the fields in the shell are required to satisfy the boundary condition $\hat{\mathbf{n}} \cdot \mathbf{j} = 0$ at both at the inner and outer interfaces of the shell ($r = R_2^-$ and $r = R_1^+$). Enforcing these boundary conditions, one obtains a reduced 4×4 homogeneous system, $\mathbf{M}_S \cdot \mathbf{v} = 0$, which can have nontrivial solutions only when $D_S \equiv \det(\mathbf{M}_S) = 0$ [48]. Here, $D_S = D_S(\omega, R_1, R_2, \omega_p, \beta)$ depends only on the parameters of the shell. The solutions in $\omega$ of $D_S = 0$ give the allowed values for the oscillation frequency of an embedded eigenstate of a given plasmonic shell. For each solution of $D_S = 0$, we introduce a transverse wave admittance $Y_w^+$ that relates the tangential electromagnetic fields at the inner interface $r = R_1^+$ as $Y_w^+ \hat{\mathbf{r}} \times \mathbf{E} = \hat{\mathbf{r}} \times (\mathbf{H} \times \hat{\mathbf{r}})$ [48]. The transverse admittance is purely imaginary $Y_w^+ = -iB_w$. On the other hand, the wave admittance $Y_w^-$ at the core side of the inner interface $r = R_1^-$ can be expressed in terms of the core parameters $R_1, \varepsilon_1$ and of $\omega$. An embedded eigenvalue can be formed only when the conditions $D_S = 0$ and $Y_w^+ = Y_w^-(\omega, R_1, \varepsilon_1)$ are simultaneously satisfied. For any solution $\omega$ of $D_S = 0$ one can generate embedded eigenstates by solving $Y_w^+ = Y_w^-(\omega, R_1, \varepsilon_1)$ with respect to the core permittivity $\varepsilon_1$. Detailed expressions for $D_S$ and $Y_w^{\pm}$ can be found in [48]. It is underlined that $D_S$ and $Y_w^{+}$ depend exclusively on the shell parameters.

Remarkably, it turns out that in the lossless limit the solutions of the reduced equation $D_S = 0$ consist of an *infinite* number of branches $\omega = \omega_{\text{trap}}^{(i)}(R_1, R_2, \omega_p, \beta)$, $i$=1,2,3… (see Fig. 1(b)). This implies that a nonlocal plasmonic shell with a given geometry may support *multiple* embedded eigenstates, rather than a unique bound state as in the local case [26]. Each solution of $D_S = 0$ corresponds to a certain surface admittance at the core interface ($Y_w^{+}$). As mentioned above, the core permittivity $\varepsilon_1$ needs to be precisely tuned to ensure that $Y_w^{+} = Y_w^{-}(\omega, R_1, \varepsilon_1)$. The insets of Fig. 1(b) show the values of $(\varepsilon_1, \varepsilon_{2,T})$ for the first three allowed eigenfrequencies. We choose solutions characterized by $\varepsilon_1 \geq 1$ (there are multiple solutions for $\varepsilon_1$ both in the local and in the nonlocal cases). The multiplicity of eigenfrequencies is a consequence of the extra degrees of freedom provided by the nonlocal response and gives the opportunity to trap light at frequencies considerably far from $\omega_p$. Different from the local case, when $\beta \neq 0$ the condition $\omega = \omega_{\text{trap}}^{(i)}$ does not lead to a zero permittivity, i.e., $\varepsilon_{2,T} \neq 0$ (the longitudinal permittivity is also nontrivial due to the wave-vector dependence).

Figures 1(c)-(d) depict the numerically calculated oscillation frequency ($\omega_{\text{trap}}$) and the corresponding wave susceptance ($B_w$) for the first three branches of solutions and for fields with a dipolar-type structure ($\text{TM}_n^r$ mode with $n$=1), as a function of the different geometrical and material parameters of the meta-atom. Figure 1(c) shows that for a strong spatial dispersion (small values of $c/\beta$), $\omega_{\text{trap}}$ and $B_w$ may differ considerably from the corresponding local values $\omega_p$ and $0$. Note that in the local regime, one has $Y_w^{+} \equiv 0$, which corresponds to a PMC boundary. As the nonlocality strength decreases ($c/\beta \to \infty$), and thereby the response of the plasmonic shell becomes increasingly local, the frequency of oscillation of the embedded eigenstate approaches $\omega_p$. The sign of $B_w$ alternates from branch to branch and is positive for the first and third branches (solid and dot-dashed curves) and negative for the second branch (dashed curve). The deviations from the local case are more significant for the higher order solution branches. Figure 1(d) illustrates the variation of $\omega_{trap}$ and $B_w$ with the normalized shell radius $R_{21} = R_2/R_1$, for a fixed value of the nonlocality strength, $\beta/c = 1/10^{3/2} = 1/31.62$ (we use a large value of $\beta/c$ to illustrate more clearly the impact of the spatial dispersion). The nonlocal effects are stronger, i.e., the frequency detuning $\omega_{trap} - \omega_p$ is larger, when the plasmonic shell is thinner.

For a specific design example, we pick $\varepsilon_1 = 1$, $R_{21} = 1.1$ and $\beta/c = 1/10^{3/2}$ and solve $D = 0$ with respect to $(\omega, R_1)$ real-valued. We obtain $R_{1,trap} = 0.973 R_{1,0}$ and $\omega_{trap} = 1.026\omega_p$ in the 1$^{st}$ branch of solutions (1$^{st}$ zero of Fig. 1(b)). The quality factor of this meta-atom is depicted in Fig. 2(a) for a detuned core radius and different values of the shell material loss. Similar to the local problem [26], for a tuned resonator ($R_1 = R_{1,trap}$) the quality factor diverges to infinity ($Q \to \infty$) when the material loss is suppressed ($\omega_c \to 0$), but in the nonlocal case for a frequency $\omega = \omega_{trap} \neq \omega_p$. Evidently, when the material response is dissipative, the quality factor and the oscillation lifetime are finite. Figure 2(b) illustrates how the complex resonance frequency $\omega = \omega' + i\omega''$ varies with the core radius for a lossless material. As expected, when the core radius matches $R_{1,trap}$ the oscillation frequency becomes real-valued, and the radiation loss is fully suppressed.

Figure 3(a) shows the electromagnetic field distribution (solid lines) of the embedded eigenstate with $R_{1,trap} = 0.973 R_{1,0}$ and $\omega_{trap} = 1.026\omega_p$. The dashed lines represent the profile of the embedded eigenstate in a meta-atom without spatial dispersion ($\omega_{trap} = \omega_p$ and $R_{1,trap} = R_{1,0}$). From Fig. 3(a), one can see that the electron-electron interactions in the plasmonic shell affect weakly the electromagnetic field distributions of the trapped field in the core region ($r < R_1$). In contrast, the fields in the plasmonic shell ($1 < r/R_1 < 1.1$) are strongly perturbed by the nonlocality. Most strikingly, the radial component of the electric field in the shell (see the curve $|E_r|_{y=0}$ in Fig. 3(a)) becomes continuous at the boundaries because the charge diffusion effects prevent the localization of a surface charge density at the interfaces. Both the local and the nonlocal models predict a strong enhancement of the radial electric field in the plasmonic shell, which is a clear fingerprint of the excitation of volume-plasmon-type oscillations. Furthermore, Fig. 3(b) reveals that the magnetic field in the spatially dispersive shell, albeit small is nontrivial. Thus, the embedded eigenstate results from the hybridization of transverse (with $\nabla \times \mathbf{E} \neq 0$) and longitudinal (with $\nabla \times \mathbf{E} = 0$) waves in the shell. Quite differently, in the local case the embedded eigenstate has a vanishing magnetic field in the shell and hence is purely longitudinal ($\nabla \times \mathbf{E} = 0$) [26]. Figure 3(d) shows time snapshots of the radial $E_r(t=0)$ electric field and of the z-component $H_z(t=0)$ of the magnetic field obtained using the nonlocal model. The dipolar structure of the field in the core is evident; the electric dipole moment is oriented along *x* and the fields have symmetry of revolution around the *x*-axis. Note that due to the symmetry of the system the eigenmode is triply degenerate.

Figure 3(c) depicts the amplitude of the embedded eigenstate fields in the plasmonic shell center as a function of $c/\beta$; the values of $R_{1,trap}$ and $\omega_{trap}$ are recalculated for each $c/\beta$. Clearly, as the nonlocality strength increases (smaller values of $c/\beta$), the amplitudes of the magnetic field and of the radial electric field in the shell are enhanced. For $\beta \to 0$ the magnetic field in the shell approaches zero.

Figures 4(a) and (b) show the electromagnetic field profiles in the shell for the second and third solutions of Fig. 1(b), respectively. The electric field profiles of the higher order modes are characterized by an increased number of maxima and nulls as compared to the first (fundamental) mode shown in Fig. 3(a). The fields in the core are similar to those of the fundamental mode (not shown).

To study the electromagnetic response of the core-shell particle under an external excitation, we consider the problem of plane wave scattering with the electric field linearly polarized. The meta-atom parameters are as in Fig. 3(a). Figure 5(a) depicts the absolute value of the Mie coefficient in the core region $\left|a_1^{TM}\right|$ as a function of frequency and for three different values of the core radius $R_1$ [48]. When $R_1$ is detuned from the optimal value $R_{1,\text{trap}}$, the Mie coefficient has a resonant behavior with a Fano-type line shape. In contrast, when $R_1$ exactly matches $R_{1,\text{trap}}$, $\left|a_1^{TM}\right| \approx 1$ has no resonant features due to a pole-zero cancellation rooted in the reciprocity of the system [26]. The reciprocity constraint can be circumvented with a nonlinear material response [29]. Specifically, with a nonlinearity the embedded eigenstates can be pumped from the outside, ensuring that the energy stored in the resonator is precisely quantized [29] (see also [18-21]). Figure 5(b) shows that the Mie coefficient in the air region $c_1^{TM}$ has a behavior analogous to $a_1^{TM}$.

The emergence of embedded eigenstates in plasmonic nanostructures is a quite unique effect. Indeed, it is fundamentally impossible to localize light in any spatially bounded (inhomogeneous) structure formed by transparent *local* isotropic dielectrics with $\varepsilon \neq 0$ and $\mu \neq 0$ [26]. To unveil the reason why spatially dispersive materials are less constrained than local materials, next we analyze the structure of the fields in the shell [26,37]. Here, we focus on layered spherical structures, but a general argument is presented in the Supplemental Material [48].

The key point is that the electromagnetic field of a $TM^r$ mode in the nonlocal shell is a superposition of two counter-propagating transverse waves and two counter-propagating longitudinal waves; hence, for a given spherical harmonic order there are $2+2=4$ degrees of freedom. In order that the radiation loss is suppressed, the electromagnetic fields outside the core-shell nanoparticle must vanish. Thus, both the tangential electromagnetic fields and the normal component of the electric current $(\hat{\mathbf{n}} \cdot \mathbf{j})$ must vanish at the shell outer interface, which corresponds to $1+1+1=3$ scalar homogeneous boundary conditions. Evidently, there is a remaining degree of freedom $(4-3=1)$, and thereby the homogeneous boundary conditions at the outer interface do not automatically force $(\mathbf{E}, \mathbf{H})$ to vanish in the shell when $\omega \neq \omega_\mathrm{p}$. In contrast, in the local limit there are only 2 degrees of freedom associated with the $TM^r$ transverse waves. In this case, the boundary conditions at the outer shell interface require the continuity of the tangential components of the fields, which correspond to 2 scalar equations. For homogeneous boundary conditions there are no extra degrees of freedom, and thus, in the local case, it is fundamentally impossible to have embedded eigenstates with $\varepsilon_2 \neq 0$ [26]. Clearly, the spatially dispersive response strongly relaxes the conditions under which the embedded eigenstates can be formed and does not require the material response to be singular in any manner.

In summary, we theoretically demonstrated that multiple embedded eigenstates with suppressed radiation loss may be supported by open spatially-dispersive core-shell meta-atoms. Surprisingly, the nonlocal effects due to electron-electron repulsive interactions do not prevent the emergence of bound states in the continuum. They rather act to strongly relax the material and geometrical conditions required for the formation of light oscillations with infinite lifetimes. Remarkably, the nonlocality enables the same material shell to perfectly screen multiple frequencies. Moreover, the material parameters of the shell do not exhibit any type of singularity. The effect is not restricted to spherical geometries, but can occur in any plasmonic resonator with two or more disjoint interfaces. Even though realistic material loss remains a practical obstacle, in principle it can be compensated using some gain mechanism [49, 50]. Thus, we believe that spatial-dispersion may provide an exciting and novel path for the realization of nanostructures with embedded eigenstates, which can have applications in optical memories and others. Furthermore, our work unveils a novel mechanism to couple radiation with matter without any form of radiation leakage.

**Acknowledgements:** This work is supported in part by the IET under the Harvey Engineering Research Prize and by Fundação para Ciência e a Tecnologia (FCT) under projects PTDC/EEITEL/ 4543/2014 and UID/EEA/50008/2019. Solange V. Silva acknowledges financial support by Fundação para a Ciência e a Tecnologia (FCT/POPH) and the cofinancing of Fundo Social Europeu under the PhD fellowship SFRH/BD/105625/2015. T. A. Morgado acknowledges financial support by FCT under the CEEC Individual 2017 contract as assistant researcher with reference CT/Nº004/2019-F00069 established with IT – Coimbra.

# Figures

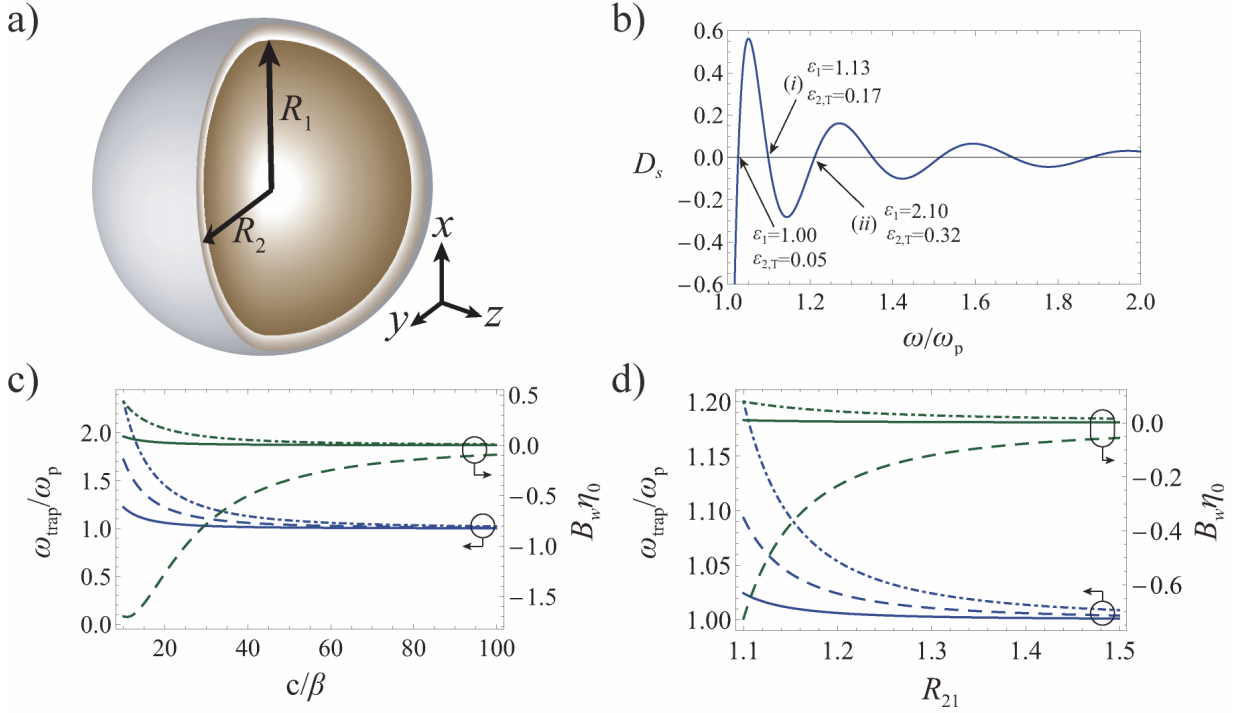

**Fig. 1.** (a) Geometry of the open bounded bi-layered spherical meta-atom. The shell is made of a plasmonic spatially-dispersive material. (b) Characteristic function $D_s$ as a function of frequency, for a meta-atom with $\beta/c = 1/10^{3/2}$, $R_{21} = 1.1$, $R_1 = 0.973 R_{10}$ and $\omega_c = 0$. The zeros indicate the values of the frequencies $\omega_{trap}^{(i)}$. The insets give the values of the core permittivity $\varepsilon_1$ and of the shell transverse permittivity $\varepsilon_{2,T}$ for the first three solutions. (c)-(d) Embedded eigenstate frequency (solid blue lines) and susceptance at the core interface (dashed green lines) as a function of the (c) locality strength $c/\beta$ and (d) normalized shell radius $R_{21} = R_2/R_1$. The susceptance is normalized to the free-space impedance $\eta_0$. The solid, dashed, and dot-dashed curves correspond to the $i=1,2,3$ solution branches, respectively. The structural parameters are $\beta/c = 1/10^{3/2}$, $\omega_c = 0$, $R_{21} = 1.1$ and $R_1/R_{1,0} = 1$, except when one of the parameters is shown in the horizontal axis of a plot.

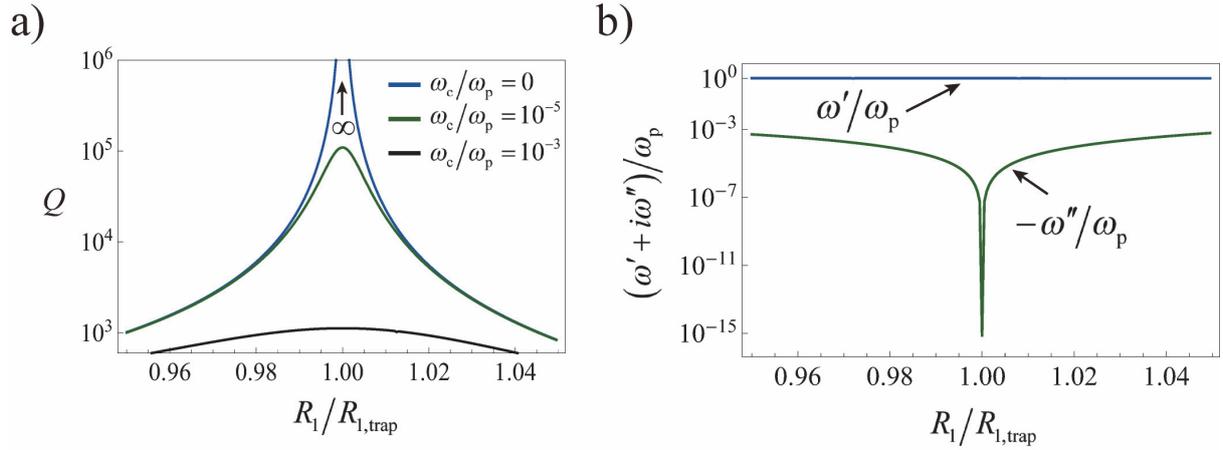

**Fig. 2.** (a) Quality factor (for the dipolar mode) as a function of the normalized core radius $R_1/R_{1,\text{trap}}$, for different values of the material loss in the plasmonic shell. (b) Variation of the real $\omega'$ and imaginary $\omega''$ parts of the eigenmode frequency $\omega = \omega' + i\omega''$ with $R_1/R_{1,\text{trap}}$, for a lossless plasmonic shell ($\omega_c/\omega_p = 0$). The structural parameters are $\beta/c = 1/10^{3/2}$, $R_{21} = 1.1$, $\varepsilon_1 = 1$. The quality factor diverges to infinity when $R_1 = R_{1,\text{trap}} = 0.973 R_{1,0}$ which yields $\omega = \omega_{\text{trap}} = 1.026\omega_p$.

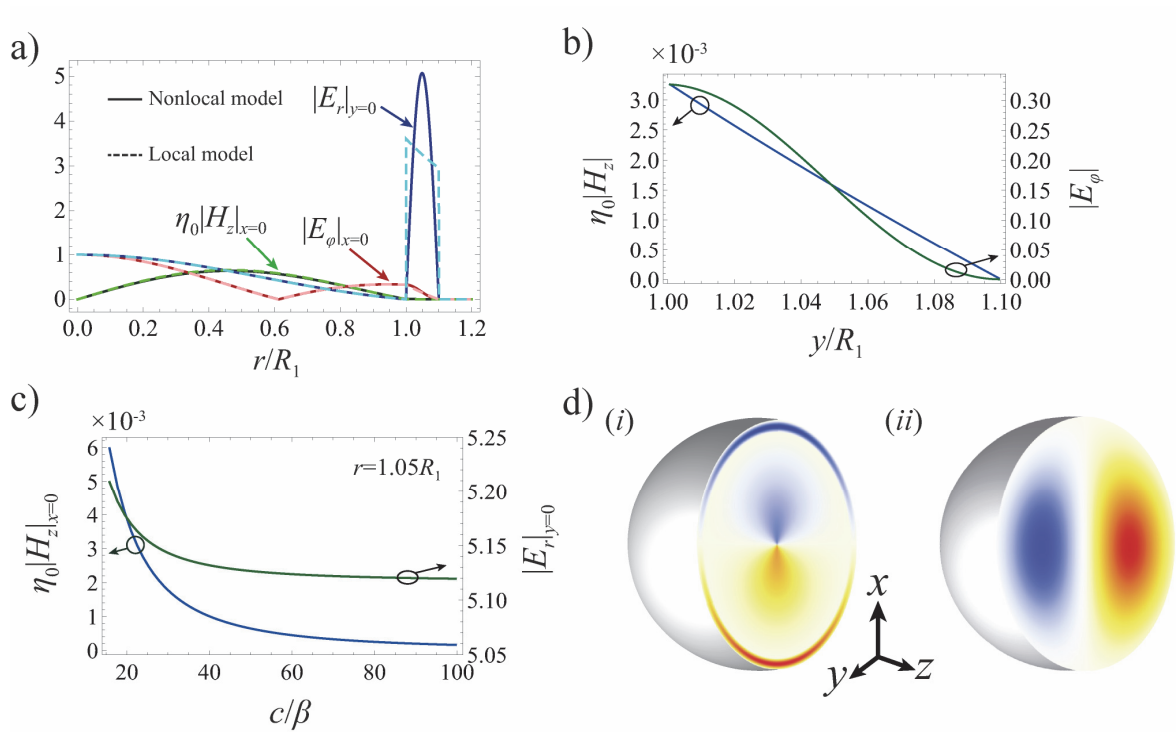

**Fig. 3.** Electromagnetic fields of an embedded eigenstate. (a) Field amplitudes in the core-shell resonator (normalized to the electric field at the core center, $r = 0$) as a function of the radial distance in the *xoy* plane, calculated using the local (dashed lines) and the nonlocal models (solid lines). Local model results: $c/\beta \to \infty$, $\omega_{\text{trap}} = \omega_p$, and $R_{1,\text{trap}} = R_{1,0}$; Nonlocal model results: $\beta/c = 1/10^{3/2}$, $\omega_{\text{trap}} = 1.026\omega_p$ and $R_{1,\text{trap}} = 0.973 R_{1,0}$. (b) Zoom of $\eta_0 |H_z|$ (blue) and $|E_\varphi|$ (green) in the shell for the nonlocal case. (c) $\eta_0 |H_z|_{x=0}$ (blue) and $|E_r|_{y=0}$ (green) as a function of $c/\beta$ at center of the shell ($r = 1.05 R_1$). (d) Time snapshots of the electric field in the nonlocal meta-atom: (*i*) $E_r(t=0)$ and (*ii*) $H_z(t=0)$ in the *xoy* plane. In all the panels, $R_{21} = 1.1$, $\varepsilon_1 = 1$, and $\omega_c = 0$.

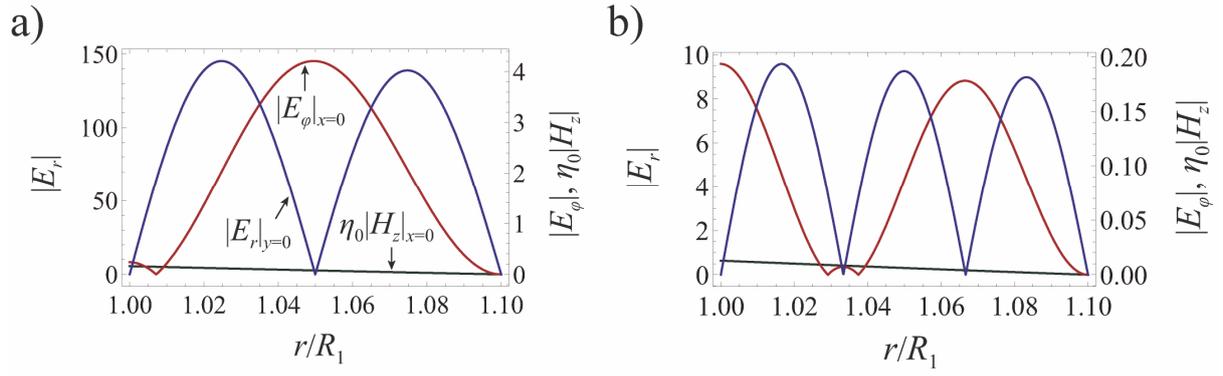

**Fig. 4.** (a)-(b) Spatial variation of the electromagnetic fields in the shell (cut in the *xoy* plane) for the modes labeled (*i*) and (*ii*) in Fig. 1b), respectively. The right vertical axis is used for $\eta_0 |H_z|_{x=0}$ and $|E_\varphi|_{x=0}$, and the left vertical axis for $|E_r|_{y=0}$.

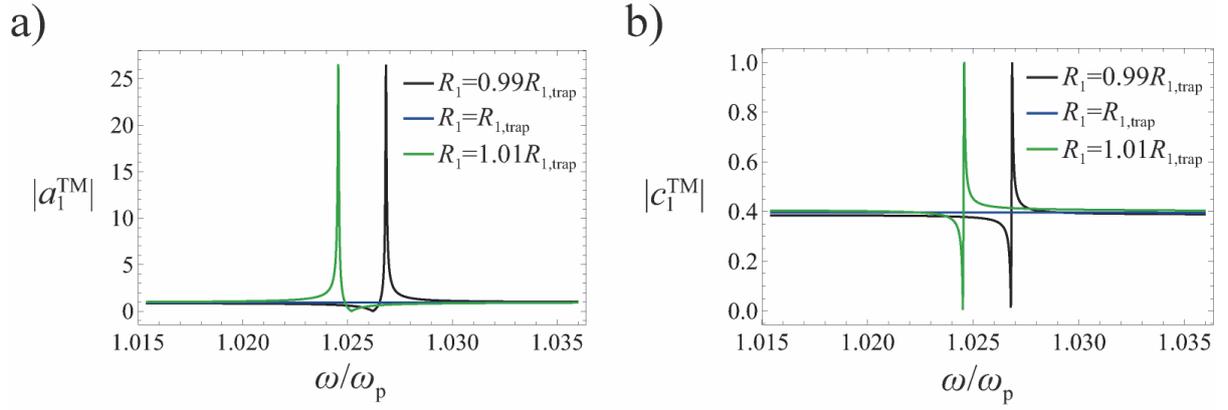

**Fig. 5.** (a) Mie coefficient $\left|a_1^{\text{TM}}\right|$ in the core region and (b) Mie coefficient $\left|c_1^{\text{TM}}\right|$ in the air region as a function of the normalized frequency $\omega/\omega_p$ and for three different values of the core radius. The meta-atom parameters are $\beta/c = 1/10^{3/2}$, $R_{21} = 1.1$, $\varepsilon_1 = 1$, and $\omega_c = 0$. The embedded eigenstate is characterized by $\omega_{\text{trap}} = 1.026\omega_p$ and $R_{1,\text{trap}} = 0.973 R_{1,0}$.

# Supplementary material for "Multiple Embedded Eigenstates in Nonlocal Plasmonic Nanostructures"

Solange V. Silva[1], Tiago A. Morgado[1], Mário G. Silveirinha[1,2*]

[1]*Instituto de Telecomunicações and Department of Electrical Engineering, University of Coimbra, 3030-290 Coimbra, Portugal*

[2]*University of Lisbon, Instituto Superior Técnico, Avenida Rovisco Pais, 1, 1049-001 Lisboa, Portugal*

*E-mail:* solange@co.it.pt, tiago.morgado@co.it.pt, mario.silveirinha@co.it.pt

Here, we derive *A*) the dispersion equation of the embedded eigenstates, *B*) the system of equations used to calculate the Mie scattering coefficients, and *C*) theoretically explain why the spatial dispersion effects enable the formation of embedded eigenstates when the shell permittivity is different from zero.

## *A. Natural modes of the spatially dispersive plasmonic core-shell nanoparticle*

Next, we obtain the characteristic equation for the natural modes of oscillation of the spatially dispersive plasmonic core-shell nanostructure (see Fig. 1(a) of the main text). Our analysis relies on the hydrodynamic (or drift-diffusion) model that takes into account the nonlocal effects in the plasmonic shell. The hydrodynamic model couples the Maxwell's equations and the continuity equation

$$\nabla \times \mathbf{E} = -\mu_0 \partial_t \mathbf{H}, \qquad \nabla \times \mathbf{H} = \mathbf{j} + \varepsilon_0 \partial_t \mathbf{E}, \tag{S1a}$$

$$\partial_t \rho = -\nabla \cdot \mathbf{j}. \tag{S1b}$$

with the Navier-Stokes equation that governs the electron transport

$$\partial_t \mathbf{j} = \varepsilon_0 \omega_p^2 \mathbf{E} - \omega_c \mathbf{j} - \beta^2 \nabla \rho. \tag{S1c}$$

---

In the above, $\rho$ is the charge density and $\mathbf{j}$ the current density. In the spectral domain ($\partial_t = -i\omega$ and $\nabla = i\mathbf{k}$) the current density $\mathbf{j}$ can be expressed in terms of the electric field. The effective permittivity of the electron gas ($\bar{\varepsilon}$) is defined such that $\mathbf{j} - i\omega\varepsilon_0 \mathbf{E} = -i\omega\varepsilon_0 \bar{\varepsilon} \cdot \mathbf{E}$. As is well-known, it is of the form:

$$\bar{\varepsilon} = \varepsilon_T \left(\mathbf{1} - \frac{1}{k^2} \mathbf{k} \otimes \mathbf{k}\right) + \varepsilon_L \frac{1}{k^2} \mathbf{k} \otimes \mathbf{k}, \tag{S2a}$$

$$\varepsilon_T(\omega) = 1 - \frac{\omega_p^2}{\omega(\omega + i\omega_c)}, \qquad \varepsilon_L(\mathbf{k}, \omega) = 1 - \frac{\omega_p^2}{\omega(\omega + i\omega_c) - \beta^2 k^2}. \tag{S2b}$$

The transverse permittivity $\varepsilon_T$ follows the Drude-type dispersion model, with $\omega_p$ the plasma frequency and $\omega_c$ the collision frequency. The longitudinal permittivity $\varepsilon_L$ depends explicity on the wave-vector ($\nabla = i\mathbf{k}$). The dispersion of the longitudinal waves is determined by

$$k^2 = \frac{1}{\beta^2}\left[\omega(\omega + i\omega_c) - \omega_p^2\right].$$

For the geometry of the main text and $TM_n^r$-polarized waves, the electromagnetic field in the plasmonic shell is a superposition of the transverse and longitudinal waves such that [R1, R2]

$$\mathbf{E} = \mathbf{E}_T + \mathbf{E}_L = \nabla \times \nabla \times \{\mathbf{r}\Psi_T(r) Y_n(\hat{\mathbf{r}})\} + \nabla \{\Psi_L(r) Y_n(\hat{\mathbf{r}})\}, \tag{S3}$$

$$\mathbf{H} = +i\omega\varepsilon_0\varepsilon_T \Psi_T(r) \hat{\mathbf{r}} \times \text{Grad } Y_n(\hat{\mathbf{r}}), \tag{S4}$$

where $Y_n$ is a spherical harmonic of order $n$ and Grad is the surface gradient operator. The functions $\Psi_T$ and $\Psi_L$ are the transverse and longitudinal potentials, respectively, and satisfy a spherical Bessel equation. For natural modes of oscillation they are given by

$$\Psi_T = \begin{cases} A\, j_n(k_1 r) & , \quad r < R_1 \\ B_{1T}\, j_n(k_T r) + B_{2T}\, y_n(k_T r) & , \quad R_1 < r < R_2 \\ C\, h_n^{(1)}(k_0 r) & , \quad r > R_2 \end{cases} \tag{S5}$$

$$\Psi_L = \begin{cases} B_{1L}\, j_n(k_L r) + B_{2L}\, y_n(k_L r) & , \quad R_1 < r < R_2 \\ 0 & , \quad \text{otherwise} \end{cases}, \tag{S6}$$

with $k_0 = \omega/c$, $k_1 = \sqrt{\varepsilon_1}\,\omega/c$, $k_T = \sqrt{\varepsilon_T}\,\omega/c$ and $k_L^2 = \dfrac{1}{\beta^2}\left[\omega(\omega + i\omega_c) - \omega_p^2\right]$. The coefficients $A$, $B_{1T}$, $B_{2T}$, $B_{1L}$, $B_{2L}$ and $C$ must ensure *i)* the continuity of the tangential components of the electromagnetic field and *ii)* that there is no electric charge flow through the shell interfaces $\hat{\mathbf{n}} \cdot \mathbf{j} = 0$. The latter constraint is the so called "additional boundary condition". From these conditions, we obtain a homogeneous linear system of equations of the form $\mathbf{M} \cdot \mathbf{x} = 0$ with $\mathbf{x} = (A, B_{1T}, B_{2T}, B_{1L}, B_{2L}, C)^T$ and

$$\mathbf{M} = \begin{pmatrix} -[j_n(k_1 r) r]'_{r=R_1} & [j_n(k_T r) r]'_{r=R_1} & [y_n(k_T r) r]'_{r=R_1} & j_n(k_L R_1) & y_n(k_L R_1) & 0 \\ -\varepsilon_1 j_n(k_1 R_1) & \varepsilon_T j_n(k_T R_1) & \varepsilon_T y_n(k_T R_1) & 0 & 0 & 0 \\ 0 & q j_n(k_T R_1) & q y_n(k_T R_1) & k_L R_1 j_n'(k_L R_1) & k_L R_1 y_n'(k_L R_1) & 0 \\ 0 & q j_n(k_T R_2) & q y_n(k_T R_2) & k_L R_2 j_n'(k_L R_2) & k_L R_2 y_n'(k_L R_2) & 0 \\ 0 & [j_n(k_T r) r]'_{r=R_2} & [y_n(k_T r) r]'_{r=R_2} & j_n(k_L R_2) & y_n(k_L R_2) & -[h_n^{(1)}(k_0 r) r]'_{r=R_2} \\ 0 & \varepsilon_T j_n(k_T R_2) & \varepsilon_T y_n(k_T R_2) & 0 & 0 & -\varepsilon_0 h_n^{(1)}(k_0 R_2) \end{pmatrix} \tag{S7}$$

with $q = (1 - \varepsilon_T) n(n+1)$ [R2]. The nontrivial solutions $\omega = \omega' + i\omega''$ (with $\omega'' \leq 0$) of the characteristic equation $D(\omega, R_1, \varepsilon_1, R_2, \omega_p, \beta) \equiv \det(\mathbf{M}) = 0$ are the natural frequencies of oscillation of the system.

We also introduce a reduced characteristic system ($\mathbf{M}_S \cdot \tilde{\mathbf{x}} = 0$ with $\tilde{\mathbf{x}} = (B_{1T}, B_{2T}, B_{1L}, B_{2L})^T$) obtained by imposing that the tangential electromagnetic field components vanish at the outer interface $r = R_2^-$, and that the additional boundary condition $\hat{\mathbf{n}} \cdot \mathbf{j} = 0$ is satisfied at the inner ($r = R_1^+$) and outer ($r = R_2^-$) interfaces. The relevant matrix is:

$$\mathbf{M}_{\mathrm{S}} = \begin{pmatrix} q j_n(k_{\mathrm{T}} R_1) & q y_n(k_{\mathrm{T}} R_1) & k_{\mathrm{L}} R_1 j'_n(k_{\mathrm{L}} R_1) & k_{\mathrm{L}} R_1 y'_n(k_{\mathrm{L}} R_1) \\ q j_n(k_{\mathrm{T}} R_2) & q y_n(k_{\mathrm{T}} R_2) & k_{\mathrm{L}} R_2 j'_n(k_{\mathrm{L}} R_2) & k_{\mathrm{L}} R_2 y'_n(k_{\mathrm{L}} R_2) \\ \left[ j_n(k_{\mathrm{T}} r) r \right]'_{r=R_2} & \left[ y_n(k_{\mathrm{T}} r) r \right]'_{r=R_2} & j_n(k_{\mathrm{L}} R_2) & y_n(k_{\mathrm{L}} R_2) \\ \varepsilon_{\mathrm{T}} j_n(k_{\mathrm{T}} R_2) & \varepsilon_{\mathrm{T}} y_n(k_{\mathrm{T}} R_2) & 0 & 0 \end{pmatrix}. \quad (S8)$$

The zeros of $D_{\mathrm{S}} \equiv \det(\mathbf{M}_{\mathrm{S}}) = 0$ in $\omega$ determine the frequencies for which a given shell can support embedded eigenstates.

For each solution of $D_{\mathrm{S}} = 0$, we introduce a transverse admittance $Y_w^+$ that links the fields at the inner shell interface ($r = R_1^+$) as $Y_w^+ \hat{\mathbf{r}} \times \mathbf{E} = \hat{\mathbf{r}} \times (\mathbf{H} \times \hat{\mathbf{r}})$. Using Eqs. (S3)-(S4) it is straightforward to show that:

$$Y_w^+ = \left. \frac{i \omega \varepsilon_0 \varepsilon_{\mathrm{T}} \psi_{\mathrm{T}}(r)}{\frac{1}{r}\left\{ [r \psi_{\mathrm{T}}(r)]' + \psi_{\mathrm{L}}(r) \right\}} \right|_{r=R_1}. \quad (S9)$$

From Eqs. (S5)-(S6) one obtains the explicit formula:

$$Y_w^+ \eta_0 = i k_0 R_1 \frac{B_{1\mathrm{T}} \varepsilon_{\mathrm{T}} j_n(k_{\mathrm{T}} R_1) + B_{2\mathrm{T}} \varepsilon_{\mathrm{T}} y_n(k_{\mathrm{T}} R_1)}{B_{1\mathrm{T}} \left[ j_n(k_{\mathrm{T}} r) r \right]'_{r=R_1} + B_{2\mathrm{T}} \left[ y_n(k_{\mathrm{T}} r) r \right]'_{r=R_1} + B_{1\mathrm{L}} j_n(k_{\mathrm{L}} R_1) + B_{2\mathrm{L}} y_n(k_{\mathrm{L}} R_1)}. \quad (S10)$$

In the above, $\eta_0$ is the free-space impedance and $(B_{1\mathrm{T}}, B_{2\mathrm{T}}, B_{1\mathrm{L}}, B_{2\mathrm{L}})^T$ is determined by the null space of $\mathbf{M}_{\mathrm{S}}$.

On the other hand, from Eq. (S5), the transverse admittance calculated at the core side of the interface ($r = R_1^-$) is:

$$Y_w^- \eta_0 = ik_0 R_1 \frac{\varepsilon_1 j_n(k_1 R_1)}{\left[r j_n(k_1 r)\right]'_{r=R_1}}. \tag{S11}$$

A solution of the reduced equation $D_S = 0$ yields an embedded eigenstate when the core permittivity $\varepsilon_1$ is tuned to ensure that $Y_w^+ = Y_w^-$. In general, there are multiple allowed solutions for $\varepsilon_1$.

Figure S1 shows how $\omega_{trap}$ and $B_w = iY_w^+$ vary with core radius $R_1$ for the same resonator parameters as in Fig. 1 of the main text. The results reveal that the frequency detuning $\omega_{trap} - \omega_p$ is larger, when the meta-atom is smaller.

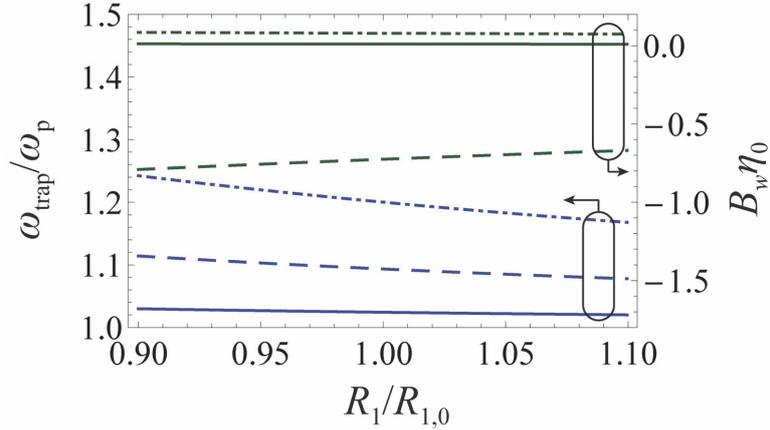

**Fig. S1.** Embedded eigenstate frequency (solid blue lines) and susceptance (dashed green lines) as a function of the normalized core radius $R_1/R_{1,0}$. The solid, dashed, and dot-dashed curves are associated with the solution branches $i=1,2,3$, respectively. The structural parameters are $\beta/c = 1/10^{3/2}$, $R_{21} = 1.1$ and the material absorption is neglected.

## *B. Scattering by a spatially dispersive plasmonic core-shell nanoparticle*

The Mie scattering coefficients for plane wave incidence can be found by expanding the fields in spherical harmonics, similar to the previous section. To take into account the incident wave, the function $\Psi_T$ must be modified as

$$\Psi_T = \begin{cases} A\, j_n(k_1 r) & , \quad r < R_1 \\ B_{1T}\, j_n(k_T r) + B_{2T}\, y_n(k_T r) & , \quad R_1 < r < R_2 \\ C\, h_n^{(1)}(k_0 r) + \dfrac{1}{ik_0} j_n(k_0 r) & , \quad r > R_2 \end{cases} \qquad (S12)$$

In the above, $a_n \to ik_1 A$ and $c_n \to ik_0 C$ are the standard Mie coefficients in the core and the air regions, respectively. Applying the same boundary conditions as in Sect. A, one obtains a linear system of equations of the form $\mathbf{M}\cdot\mathbf{x}=\mathbf{b}$, where $\mathbf{M}$ is given by the matrix (S7) and $\mathbf{b}$ is the vector:

$$\mathbf{b} = \frac{1}{ik_0}\left( 0 \quad 0 \quad 0 \quad 0 \quad \left[j_n(k_0 r)r\right]'_{r=R_2} \quad j_n(k_0 R_2) \right)^T. \qquad (S13)$$

## *C. Justification why a nonlocal shell can support embedded eigenstates*

In the lossless limit, the hydrodynamic model [Eqs. (S1a)-(S1c)] can be written in a compact form as

$$\mathbf{M}_g \cdot i\partial_t \mathbf{Q} = \hat{L}\cdot\mathbf{Q}, \qquad (S14)$$

where $\mathbf{Q}=(\mathbf{E},\mathbf{H},\tilde{\mathbf{j}},\tilde{\rho})^T$ represents a ten-component state vector with $\tilde{\mathbf{j}}=\mathbf{j}/\sqrt{\varepsilon_0 \omega_p^2}$ and $\tilde{\rho}=\rho\beta/\sqrt{\varepsilon_0 \omega_p^2}$. Here, $\hat{L}$ is a first-order linear-differential Hermitian operator defined as

$$\hat{L} = \begin{pmatrix} \mathbf{0} & i\nabla\times\mathbf{1} & -i\sqrt{\varepsilon_0 \omega_p^2}\mathbf{1} & 0 \\ -i\nabla\times\mathbf{1} & \mathbf{0} & \mathbf{0} & 0 \\ i\sqrt{\varepsilon_0 \omega_p^2}\mathbf{1} & \mathbf{0} & \mathbf{0} & -i\beta\nabla \\ 0 & 0 & -i\beta\nabla\cdot & 0 \end{pmatrix}, \qquad (S15)$$

where $\mathbf{1}$ is the 3x3 identity matrix and $\mathbf{0}$ is a 3x3 matrix filled with zeros, and $\mathbf{M}_g$ is a material matrix given by

$$\mathbf{M}_g = \begin{pmatrix} \varepsilon_0 \mathbf{1} & 0 & 0 & 0 \\ 0 & \mu_0 \mathbf{1} & 0 & 0 \\ 0 & 0 & 1 & 0 \\ 0 & 0 & 0 & 1 \end{pmatrix}. \tag{S16}$$

Let us introduce the Green's function $\mathbf{G}$ (for a homogeneous unbounded space) that satisfies $\hat{L} \cdot \mathbf{G} = \omega \mathbf{M}_g \cdot \mathbf{G} + i\delta(\mathbf{r} - \mathbf{r}')\mathbf{1}_g$. Consider now some solution of $\hat{L} \cdot \mathbf{Q} = \omega \mathbf{M}_g \cdot \mathbf{Q}$ defined in the interior of some volumetric region of space $V$ (the plasmonic shell). Let $\Sigma$ stand for the boundary surface enclosing volume $V$. The considered function can be trivially extended to all space (with the state vector identical to zero outside $V$) as the solution of:

$$\hat{L} \cdot \mathbf{Q} = \omega \mathbf{M}_g \cdot \mathbf{Q} + \begin{pmatrix} -i\hat{\mathbf{n}} \times \mathbf{H} \\ i\hat{\mathbf{n}} \times \mathbf{E} \\ \beta i \hat{\mathbf{n}} \tilde{\rho} \\ i\beta \hat{\mathbf{n}} \cdot \tilde{\mathbf{j}} \end{pmatrix} \delta_\Sigma, \tag{S17}$$

where $\delta_\Sigma$ represents a delta-function type distribution that vanishes outside $\Sigma$, which when integrated over all space gives the area of $\Sigma$; $\hat{\mathbf{n}}$ is the outward unity normal vector. From the definition of the Green function, we have

$$\mathbf{Q}(\mathbf{r}) = \int_\Sigma ds' \, \mathbf{G}(\mathbf{r} - \mathbf{r}') \cdot \begin{pmatrix} -\hat{\mathbf{n}} \times \mathbf{H}(\mathbf{r}') \\ \hat{\mathbf{n}} \times \mathbf{E}(\mathbf{r}') \\ \beta \hat{\mathbf{n}} \tilde{\rho}(\mathbf{r}') \\ \beta \hat{\mathbf{n}} \cdot \tilde{\mathbf{j}}(\mathbf{r}') \end{pmatrix}. \tag{S18}$$

Let us consider that $\Sigma$ (the shell boundary) is formed by external and internal surfaces $\Sigma = \Sigma_{out} \cup \Sigma_{int}$. Suppose also that the tangential electromagnetic fields and the normal component of the current vanish at the outer shell boundary, which are the necessary conditions for the formation of an embedded eigenstate. Then, Eq. (S18) becomes (imposing also that the normal component of the current vanishes at the inner shell interface)

$$\mathbf{Q}(\mathbf{r}) = \int_{\Sigma_{\text{out}}} ds' \, \mathbf{G}(\mathbf{r}-\mathbf{r}') \cdot \begin{pmatrix} \mathbf{0} \\ \mathbf{0} \\ \beta \hat{\mathbf{n}} \tilde{\rho}(\mathbf{r}') \\ 0 \end{pmatrix} + \int_{\Sigma_{\text{int}}} ds' \, \mathbf{G}(\mathbf{r}-\mathbf{r}') \cdot \begin{pmatrix} -\hat{\mathbf{n}} \times \mathbf{H}(\mathbf{r}') \\ \hat{\mathbf{n}} \times \mathbf{E}(\mathbf{r}') \\ \beta \hat{\mathbf{n}} \tilde{\rho}(\mathbf{r}') \\ 0 \end{pmatrix}. \quad \text{(S19)}$$

Importantly, Eq. (S19) shows that notwithstanding the homogeneous boundary conditions on the outer surface, the corresponding surface integral over $\Sigma_{\text{out}}$ is not suppressed because $\tilde{\rho}|_{\Sigma_{\text{out}}}$ can be nontrivial. Due to this reason, different from the local case discussed in [R2], it is not feasible to use analytical continuation arguments to conclude that $\mathbf{Q}$ vanishes inside the shell (volume $V$). In other words, for layered structures, the state vector $\mathbf{Q}$ always has a contribution from the outer surface, and thereby does not need to vanish in the shell region. This explains why embedded eigenstates can be formed in a wide spectral range when the nonlocality of the shell is taken into account.

Furthermore, Eq. (S19) can be understood as an homogeneous integral equation with respect to the unknowns $\tilde{\rho}|_{\Sigma_{\text{out}}}$, $\tilde{\rho}|_{\Sigma_{\text{in}}}$, $\hat{\mathbf{n}} \times \mathbf{E}|_{\Sigma_{\text{in}}}$ and $\hat{\mathbf{n}} \times \mathbf{H}|_{\Sigma_{\text{in}}}$ and subject to the constraints $\hat{\mathbf{n}} \times \mathbf{E}|_{\Sigma_{\text{out}}} = 0$ and $\hat{\mathbf{n}} \times \mathbf{H}|_{\Sigma_{\text{out}}} = 0$, $\hat{\mathbf{n}} \cdot \tilde{\mathbf{j}}|_{\Sigma_{\text{out}}} = 0$ and $\hat{\mathbf{n}} \cdot \tilde{\mathbf{j}}|_{\Sigma_{\text{in}}} = 0$, This homogenous integral equation generalizes the reduced characteristic system discussed in Sect. A. The homogeneous integral equation can have non-trivial solutions only for specific values of $\omega$ which depend exclusively on the geometrical shape of the plasmonic resonator, and on $\omega_{\text{p}}, \beta$. Such values of $\omega$ determine the frequencies of the allowed embedded eigenstates.